# Magnetization reversal and negative volume thermal expansion in Fe doped $Ca_2RuO_4$

S. J. Yuan[1*], T. F. Qi[1], J. Terzic[1], Hao Zheng[1], Zhao Zhao[2], Songxue Chi[3], Feng Ye[3], Hua Wei[4], S. Parkin[5], Xuerong Liu[6,7], Wendy L. Mao[8,9], G. Cao[1*]

*[1] Center for Advanced Materials and Department of Physics and Astronomy, University of Kentucky, Lexington, KY 40506, USA*

*[2]Department of Physics, Stanford University, Stanford, California 94305, USA*

*[3]Quantum Condensed Matter Division, Oak Ridge National Laboratory, Oak Ridge, Tennessee 37831, USA*

*[4]Scintillation Materials Research Center, Department of Materials Science and Engineering, University of Tennessee, Knoxville, Tennessee 37996, USA*

*[5]Department of Chemistry, University of Kentucky, Lexington, KY 40506, USA*

*[6]Institute of Physics, Chinese Academy of Sciences, Beijing 100190, China*

*[7]Condensed Matter Physics and Materials Science Department, Brookhaven National Laboratory, Upton, New York 11973, USA*

*[8]Department of Geological Sciences, Stanford University, Stanford, CA 94305, USA*

*[9]Stanford Institute for Materials and Energy Sciences, SLAC National Accelerator Laboratory, Menlo Park, CA 94025, USA*

## ABSTRACT

We report physical and structural properties of single-crystal $Ca_2Ru_{1-x}Fe_xO_4$ ($0 \le x \le 0.20$) as functions of temperature, magnetic field and pressure. $Ca_2RuO_4$ is a structurally-driven Mott insulator with a metal-insulator (MI) transition at $T_{MI}$ = 357 K, which is well separated from antiferromagnetic order at $T_N$ = 110 K. Fe substitution for Ru in $Ca_2RuO_4$ causes a pronounced magnetization reversal and giant negative volume thermal expansion (NVTE). The magnetization reversal is a result of a field-induced antiferromagnetic coupling between the Ru- and Fe-magnetic sublattices that have different temperature dependence. The NVTE is closely associated with the orthorhombic distortion, and becomes smaller as the orthorhombicity weakens due to either Fe doping or application of pressure. The study highlights an intriguing interplay between lattice, orbital and spin degrees of freedom that is at the root of the novel phenomena in $Ca_2RuO_4$.

## I. INTRODUCTION

The layered ruthenates are among the most studied transition mental oxides in recent years chiefly because the comparable magnitudes of their intra-atomic Coulomb interaction $U$ and 4d-bandwidth W leave the ruthenates precariously balanced on the border between metallic and insulating behavior, or on the verge of long-range magnetic order [1]. Of the layered ruthenates, $Ca_2RuO_4$ has attracted particular interests owing to its peculiar interplay between crystal and electronic structures that drives exotic states [1-14]. Extensive investigations on $Ca_2RuO_4$ have established that a strong cooperative Jahn-Teller distortion removes the degeneracy of the three Ru $t_{2g}$ orbitals ($d_{xy}$, $d_{yz}$, $d_{zx}$) via a transition to an orbital order that, in turn, drives a temperature-induced metal-insulator (MI) transition at $T_{MI}$ = 357 K [2-14]. However, unlike other Mott insulators, $Ca_2RuO_4$ undergoes an antiferromagnetic (AFM) order at $T_N$ = 110 K, much lower than $T_{MI}$ [2,5], making it an archetype of a MI transition that is driven by a structural transition from a high-temperature tetragonal to low-temperature orthorhombic distortion [5,6,12-14].

We have recently observed that slight substitutions of a 3*d* transition metal element M (M = Cr, Mn, Fe and Cu) for Ru in $Ca_2RuO_4$ shifts $T_{MI}$ to lower temperature, weakens the orthorhombic distortion, and induces substantial negative volume thermal expansion (NVTE) in $Ca_2Ru_{1-x}M_xO_4$ [14-16]. In contrast, slight substitution of 5*d*-element Ir for Ru in $Ca_2RuO_4$ strengthens the orthorhombic distortion, and thus shifts $T_{MI}$ to higher temperatures with a normal positive thermal expansion [17]. The onset of the NVTE in $Ca_2Ru_{1-x}M_xO_4$ is strongly coupled with the magnetic and orbital orders, which closely tracks the changing orthorhombicity as $x$ changes. The NVTE disappears when the orthorhombicity vanishes near a critical doping concentration. These observations highlight a complex interplay between orbital, spin, and lattice degrees of freedom that is unique to $Ca_2RuO_4$. In this paper, we report results of our more recent study on Fe doped $Ca_2RuO_4$ or single-crystal $Ca_2Ru_{1-x}Fe_xO_4$ ($0 \leq x \leq 0.20$). Besides the NVTE that is induced by Fe doping, we have also observed a pronounced magnetization reversal in which the net magnetic moments are antiparallel to the applied magnetic field, *H*, an energetically unfavorable configuration under normal circumstances. We attribute this unusual phenomenon to a field-induced antiferromagnetic coupling between the magnetic moments of Ru- and Fe-sublattices that have different temperature dependence. Moreover, we present results of a detailed crystal structural study as functions of temperature and pressure (up to 10.6 GPa), which illustrate a similar effect to Fe doping, confirming the importance of the orthorhombic distortion to the NVTE.

## II. EXPERIMENT

The $Ca_2Ru_{1-x}Fe_xO_4$ ($0 \leq x \leq 0.20$) single crystals were grown using a NEC two-mirror floating zone optical furnace. Details of single-crystal growth are described elsewhere [15]. The single-crystal x-ray diffraction was performed as a function of temperature between 10 K and 430 K using a Nonius-Kappa CCD single-crystal x-ray diffractometer. Structures were refined by full-matrix least squares using the SHELX-97 programs [18]. All refinements were performed with the same settings, constraints and restraints. The structures affected by absorption were corrected by comparison of symmetry-equivalent reflections using the program SADABS[19]. It needs to be emphasized that the single crystals are of high quality and there is no indication of any mixed phases or inhomogeneity in all doped single crystals studied. Chemical compositions of the single crystals were estimated using both single-crystal x-ray diffraction and energy dispersive x-ray (EDX) analysis (Hitachi/Oxford 3000). Magnetization, specific heat, and electrical resistivity were measured using either a Quantum Design superconducting quantum interference device magnetometer and/or physical property measurement system. High pressure and high temperature powder x-ray diffraction measurements were performed at 16 BMD, Advanced Photon Source at Argonne National Laboratory. Single crystals of $Ca_2Ru_{0.92}Fe_{0.08}O_4$ were ground into fine powder sample. A tungsten foil was pre-indented up to 10 GPa and drilled with a sample chamber of 150 μm in diameter. Silicone oil was used as the pressure transmitting medium and ruby was used as the pressure calibrant. High pressure was generated in a symmetric diamond anvil cell with culet size of 500 microns. The entire diamond anvil cell assembly was heated for the high temperature measurements. Rietveld refinements were performed using the GSAS-EXPGUI package[20].

## III. RESULTS AND DISCUSSION

### A. Magnetization reversal

The temperature dependence of the magnetization in the *ab* plane $M_{ab}$(T) of the parent compound $Ca_2RuO_4$ exhibits a sharp AFM order at $T_N$ = 110 K (see **Fig.1 (a)**). The magnetic property of $Ca_2RuO_4$ is very sensitive to Fe doping and exhibits pronounced changes, as shown in **Fig. 1 (a) & (b)**. For $0 \leq x \leq 0.05$, the ground state remains robust AFM state with a slight drop in magnitude. For $x$ = 0.08, $M_{ab}$(T) shows a wide peak below $T_N$ = 110 K indicating a weak ferromagnetic (FM) moments. Most strikingly, this behavior is then evolved into a magnetization reversal for $0.09 \leq x \leq 0.20$ around 37 K. Here, $M_{ab}$(T) was measured using a field-cooling (FC) sequence, i.e., the sample was first field-cooled at $\mu_0 H$ = 0.1 T from room temperature to 1.7 K, then measured during warming process with $\mu_0 H$ = 0.1 T. For the zero-field-cooling (ZFC) sequence, $M_{ab}$(T) shows no magnetization reversal; instead, it remains positive and exhibits a sharp AFM order around $T_N$ = 110 K, distinctly different from that of the FC $M_{ab}$(T) [see inset in Fig.1(b)]. It is clear that the magnetization reversal and magnetic behavior below $T_N$ are field-induced. The magnetization reversal becomes more pronounced with increasing Fe concentration until $x$ > 0.20 where $T_N$ vanishes due to the weakened orthorhombic distortion [16], as shown in Fig.1 (b). A magnetization reversal is unusual but has been observed in other antiferromagnets, such as orthoferrites [21,22], orthochromites [23-25], orthovanadates [26,27], manganites [28,29] and iridates [30]. The Fe doping for Ru in $Ca_2RuO_4$ likely leads to two inequivalent magnetic sublattices i.e. the Ru and Fe sublattices that are antiferromagnetically coupled; the magnetization reversal could be a result of different temperature dependences of these two magnetic sublattices. During the FC process from room temperature to $T_N$, the spins of the Ru sublattice are orientated along the direction of the applied field whereas the spins of the Fe sublattice are antiparallel to the applied field. Since the Ru and Fe magnetic sublattices have different temperature dependence, the magnetic moment of the Fe sublattice dominates that of the Ru sublattice at low temperatures, resulting in the negative $M_{ab}$ [see the inset of Fig.1(b)].

The magnetization reversal required a close examination. Detailed measurements were carried out for $x$ = 0.12. The sample was first field-cooled at various magnetic fields from room temperature to 1.7 K where the field was removed before the data were collected; therefore, the magnetic data presented in Fig. 2(a) were taken at ***zero-field*** in a warming process. The absolute value of the remnant magnetization $M_{ab}$(T) increases with increasing cooling-field, indicating a field-induced behavior. The complication of this field-behavior is discussed below in detail. All the remnant magnetization $M_{ab}$(T) cooled at different magnetic fields converges at a compensation temperature $T_{com}$ = 37 K, a characteristic temperature for a ferrimagnetic state. $M_{ab}$(T) measured using a standard FC process is presented in Figure 2 (b). The magnetization reversal is suppressed with increasing $H$ and disappears when $\mu_0 H$ > 3 T, suggesting that the direction of net magnetic moment that is initially opposite to the field direction becomes aligned with the direction of the applied field at $\mu_0 H$ > 3 T.

The isothermal magnetization $M_{ab}$(H) for $x$ = 0.12 were measured at various temperatures after the ZFC procedure, as shown in Fig.2 (c). For $T_{com} < T < T_N$, $M_{ab}$(H) shows apparent magnetic loops yet with extremely low moments, indicating weak ferromagnetism or a canted AFM structure. On the other hand, for $T < T_{com}$, $M_{ab}$(H) measured using either the ZFC or FC process show linear field-dependence, indicating an antiferromagnetic configuration without canting. Moreover, there is a negative remnant moment (i.e., at $\mu_0 H$ = 0) in the FC case, which is absent in the ZFC case [see Fig2. (d)]. The profound difference is a result of the field-induced behavior: In the ZFC case, the absence of H leaves the magnetic moments of the Fe sublattice in a disordered state. Therefore, the observed magnetic behavior is primarily due to the Ru sublattice, hence, no negative remnant moment occurs [see Fig2. (d)]. However, in the FC case, the finite applied magnetic field induces an ordered state in the Fe sublattice. This ordered Fe sublattice, in turn, interlocks with the ordered Ru sublattice in an antiparallel fashion during the FC process, forming a ferrimagnetic configuration. Furthermore, spins of the Fe sublattice are likely to be antiparallel to the direction of $H$ and become dominant below $T_{com}$, leading to a negative vertical magnetization shift in $M_{ab}$(H) as shown in Fig2.

(d). $M_{ab}$(H) becomes positive when $\mu_0 H > 3$ T when spins of the Ru sublattice predominate. A remarkable feature of this magnetization reversal is that a magnetic loop that is often anticipated is conspicuously absent.

The magnetic structure of the parent compound $Ca_2RuO_4$ with a sharp AFM order at 110 K is an *A*-centered magnetic mode (1 0 0) ($La_2CuO_4$ type) [ see Fig. 10 in Ref.[4] ]. Results of a recent neutron study show that Fe doping favors a *B*-centered magnetic mode (1 0 1) ($La_2NiO_4$ type) and gradually kills the *A*-centered mode [31]. In essence, for $0 \le x \le 0.05$, the system possesses the *A*-centered mode inherent in the parent compound. With increasing *x*, e.g., at $x = 0.08$, the neutron results show that the *B*-centered mode which favors FM coupling becomes dominant around 110 K and recedes near 70 K, below which the A-centered mode starts to dominate instead [31]. This explains the weak FM moments (i.e. wide peak) in the temperature range of 70 K $\le T \le$ 110 K. As *x* further increases, the *B*-centered mode prevails over a larger temperature range [31], which explains the wider FM peak in Fig. 1(b).

### B. Negative volume thermal expansion

Fe doped $Ca_2RuO_4$ preserves the low-temperature orthorhombic symmetry (*Pbca*) but weakens the orthorhombic distortion by reducing the difference between the *a* and *b* axis and elongating the *c* axis, reported previously [see Fig. 5 in Ref.[16]. We now focus on the coupling between the NVTE and MI transition at $T_{MI}$ and the magnetic order at $T_N$ by examining two sets of representative data for $x = 0.08$ and 0.12. At $x$ =0.08, as shown in Fig.3, the right shaded area clearly illustrates that the abrupt expansion of the unit cell volume (*V*) occurs at $T_{MI}$ that is characterized by a strong anomaly in the heat capacity. There is an exception where the volume *V* contracts normally on cooling between 70-110

K, and then expands again when the AFM transition is completely established near 70 K, as shown in the left shaded area in Fig. 3. The lattice change closely follows the change of the magnetic structure between 70-110 K, where the A and B magnetic modes compete, as discussed above. This close association indicates a strong magnetoelastic effect. Below 70 K, the NVTE prevails again because of a more rigid spin configuration, in which the *A*-centered mode dominates. On the other hand, at higher Fe doping level, $x = 0.12$, the orthorhombicity is considerably weakened with the disappearance of the first-order structural transition and the MT transition. Concomitantly, the abrupt $V$ expansion at $T_{MI}$ diminishes. Instead, $V$ shows a continuous expansion yet with a smaller volume expansion ratio, $\Delta V/V \approx 0.5\%$ on cooling, as illustrated in Fig. 4 (b), in contrast with $\Delta V/V \approx 0.8\%$ for $x = 0.08$. The simultaneous disappearance of both $T_{MI}$ and the abrupt expansion of $V$ reinforces that the first-order transition of volume and the orbital order are indeed strongly coupled.

### C. High-pressure study

We carried out a high-pressure study on one representative concentration $x = 0.08$. We first investigated changes in the structural transition at a fixed pressure of 0.6 GPa in a temperature range 304 K $\leq T \leq$ 421 K. The high-pressure results together with the ambient pressure ones are documented in Fig. 5. At ambient pressure, the first-order structural transition from a high-temperature tetragonal to a low-temperature orthorhombic phase occurs abruptly at 380 K. In contrast, this structural change at 0.6 GPa appears smooth without a sign of the structural phase transition at more elevated temperatures (see Fig.5). The difference suggests that the first-order structural transition along with the abrupt volume change is suppressed under pressure. Notice that this effect is similar to that of high Fe doping in $Ca_2RuO_4$, as shown in Fig.4.

The pressure dependence of room temperature lattice parameters for $x$ = 0.08 is determined up to 10.6 GPa, and the results are shown in Fig.6. Three distinct Phases I, II and III are defined according the lattice properties. The discontinuous transition from the Phase I with a shorter *c*-lattice parameter to the Phase II with a longer *c*-lattice parameter takes place around 0.6 GPa with a dramatic volume decrease by 1.2%. With further increasing the applied pressure, the orthorhombic splitting shows the opposite sign with $a > b$ at 4.2 GPa (see Fig.6(a)), and the lattice parameters *a*, *b*, and *c*, and V display an anomaly characterized by a change of slope between 4.9 and 5.7 GPa [see Fig.6]. At pressures above 5.7 GPa, the powder XRD patterns can be equally refined by *Pbca* and *Bbcm* structures. Considering *Bbcm* has higher symmetry than *Pbca*, *Bbcm* space group is assigned to data, in accordance to previous high pressure work on $Ca_2RuO_4$ [32].

Our study suggests that applying Fe doping and high pressure on $Ca_2RuO_4$ have quite similar effects on the crystal structure. Moderate Fe doping and external pressure can effectively weaken the orthorhombic distortion and induce NVTE. Higher Fe doping ($x \geq$ 0.12) and external pressure ($\geq$ 0.6 GPa ) suppress the first-order structural transition, and thus also suppress the MI transition and the abrupt *V* expansion.

## IV. CONCLUSIONS

Fe substitution for Ru in $Ca_2RuO_4$ causes the magnetization reversal and negative volume thermal expansion. The field-induced magnetic moments in the Fe-sublattice are antiferromagnetically coupled with those in the Ru-sublattice below the compensation $T_{com}$ = 37 K; the different temperature dependence of the Fe- and Ru-sublattices results in the net magnetic moments that are antiparallel to the direction of *H*, thus the magnetization reversal in $Ca_2Ru_{1-x}Fe_xO_4$. The negative volume thermal expansion is strongly coupled to the magnetic state and structural phase transition. The brief appearance of the positive or normal thermal expansion in 70 K - 110 K is attributed to a frustrated magnetic state due to the competing A- and B-centered modes. The application of high pressure results in a similar effect of Fe doping, confirming the importance of the orthorhombic distortion to the NVTE. All results indicate a unique and delicate coupling between the crystal and magnetic structures inherent in $Ca_2RuO_4$. It is this coupling that is at the root of a rich phase diagram in the ruthenate.

**ACKNOWLEDGEMENTS**

This work was supported by the National Science Foundation via Grant No. DMR-1265162. Work by Z. Zhao and W. L. Mao was supported by the Department of Energy (DOE), Basic Energy Sciences (BES), Materials Sciences and Engineering Division, under Contract DE-AC02-76SF00515. HPCAT operations are supported by DOE-NNSA, DE-NA0001974 and DOE-BES, DE-FG02-99ER45775, with partial instrumentation funding by NSF MRI-1126249. APS is supported by DOE-BES, DE-AC02-06CH11357.

*Corresponding authors: sjyuan.shu@gmail.com; cao@uky.edu

## REFERRENCES

**FIGURE CAPTIONS:**

**Fig. 1.** The temperature dependence of magnetization in *ab* plane $M_{ab}(T)$ for $CaRu_{1-x}Fe_xO_3$, where (a) $0 \leq x \leq 0.05$ and (b) $0.08 \leq x \leq 0.20$. The data were collected after field cooling procedure at $\mu_0 H$ =0.1 T. The inset in (b) shows $M_{ab}(T)$ for $x$ = 0.12 after zero-field-cooling (ZFC) and field-cooling (FC) sequences. The arrows in the inset represent the evolution of the effective magnetic moment of the Fe (in red) and Ru (in black) ions.

**Fig. 2.** For $Ca_2Ru_{1-x}Fe_xO_4$ with $x$ = 0.12: (a) the remnant magnetization $M_{ab}(T)$ measured in zero field after FC process; (b) the $M_{ab}(T)$ after FC from room temperature down to $T$ = 1.7 K with various applied field $\mu_0 H$ = 0.1, 1, 2, 3, 4, 5, 6 and 7 T (the cooling field and measuring field are same); c) isothermal magnetization $M_{ab}$ at 5, 30, 60, 80, 100 and 120K for the ZFC sequences and (d) isothermal magnetization $M_{ab}$ at 5 K after the ZFC and FC sequences.

**Fig. 3.** For $Ca_2Ru_{1-x}Fe_xO_4$ with $x$ = 0.08, the temperature dependences of (a) lattice parameters *a*, *b*, and *c* axes (right scale); (b) unit cell volume *V* and specific heat C; and (c) magnetic susceptibility at $\mu_0 H$ = 0.1 T (field cooled) and ab plane resistivity (right scale). The left shaded area indicates the concomitant occurrence of the normal thermal expansion and weak FM characterized by a small anomaly in the specific heat $C(T)$; the right shaded area indicates the concomitant occurrence of the NVTE and MI transition characterized by a strong anomaly in the specific heat $C(T)$.

**Fig.4.** For $Ca_2Ru_{1-x}Fe_xO_4$ with $x$ = 0.12, the temperature dependences of (a) lattice parameters *a*, *b*, and *c* axes (right scale); (b) unit cell volume *V* and specific heat C; and (c) magnetic susceptibility at $\mu_0 H$ = 0.1 T (field cooled) and ab plane resistivity (right scale).

**Fig. 5.** For $Ca_2Ru_{1-x}Fe_xO_4$ with $x = 0.08$, the temperature dependence of lattice constants under ambient pressure and 0.6 GPa (a) *a* and *b* axes, (b) *c* axis and (c) unit cell volume. The solid symbols represent the lattice parameters under ambient pressure, and the open symbols represent the lattice parameters under 0.6 GPa. Errors given by GSAS-EXPGUI are smaller than size of the markers.

**Fig. 6**. For $Ca_2Ru_{1-x}Fe_xO_4$ with $x = 0.08$ at room temperature, the pressure dependence of lattice constants (a) *a* and *b* axes, (b) *c* axis and (c) unit cell volume. The solid lines are guides to show change in slope of lattice constants and volume. Three structural phases are noted in the figure and separated by vertical dashed lines.

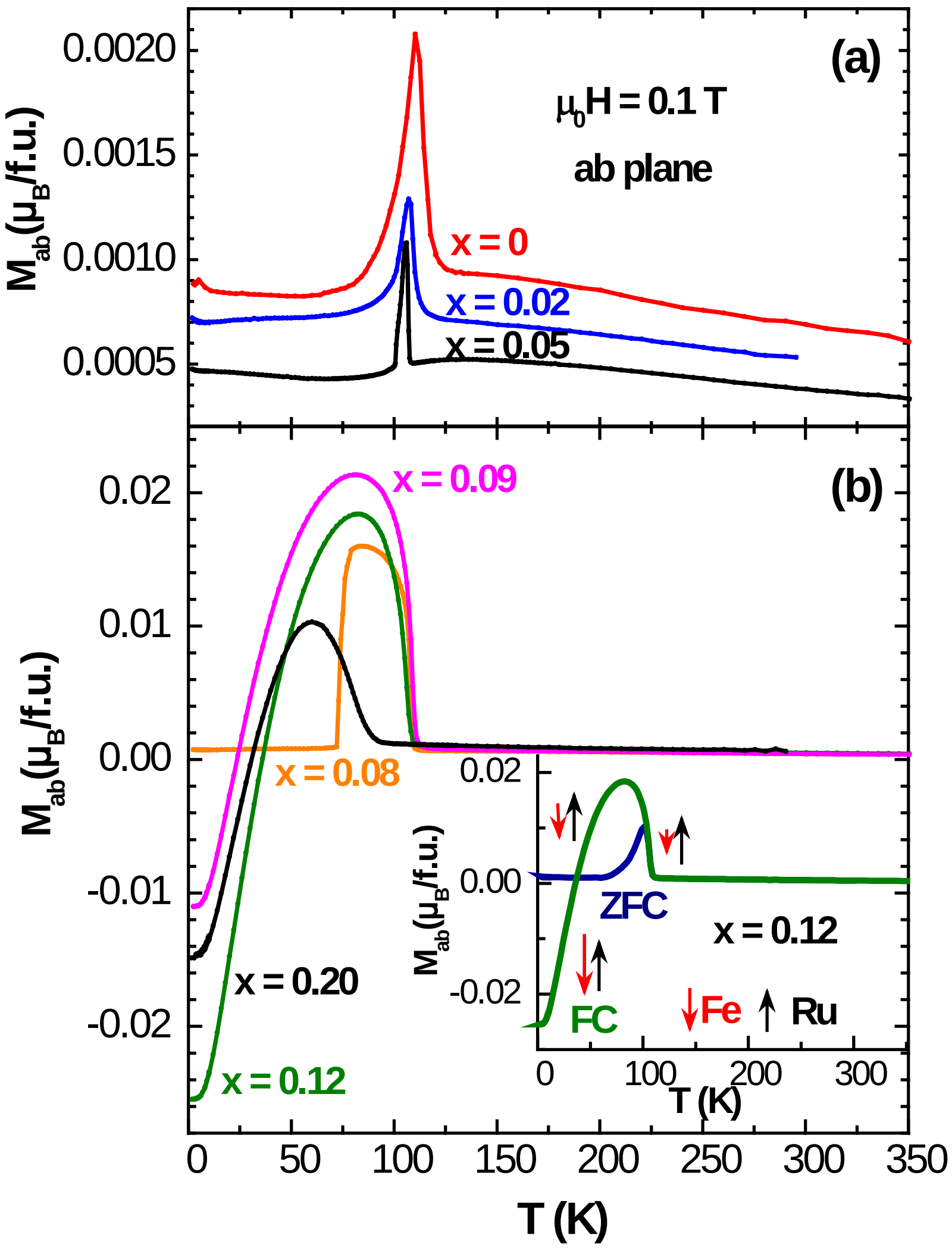

Fig. 1

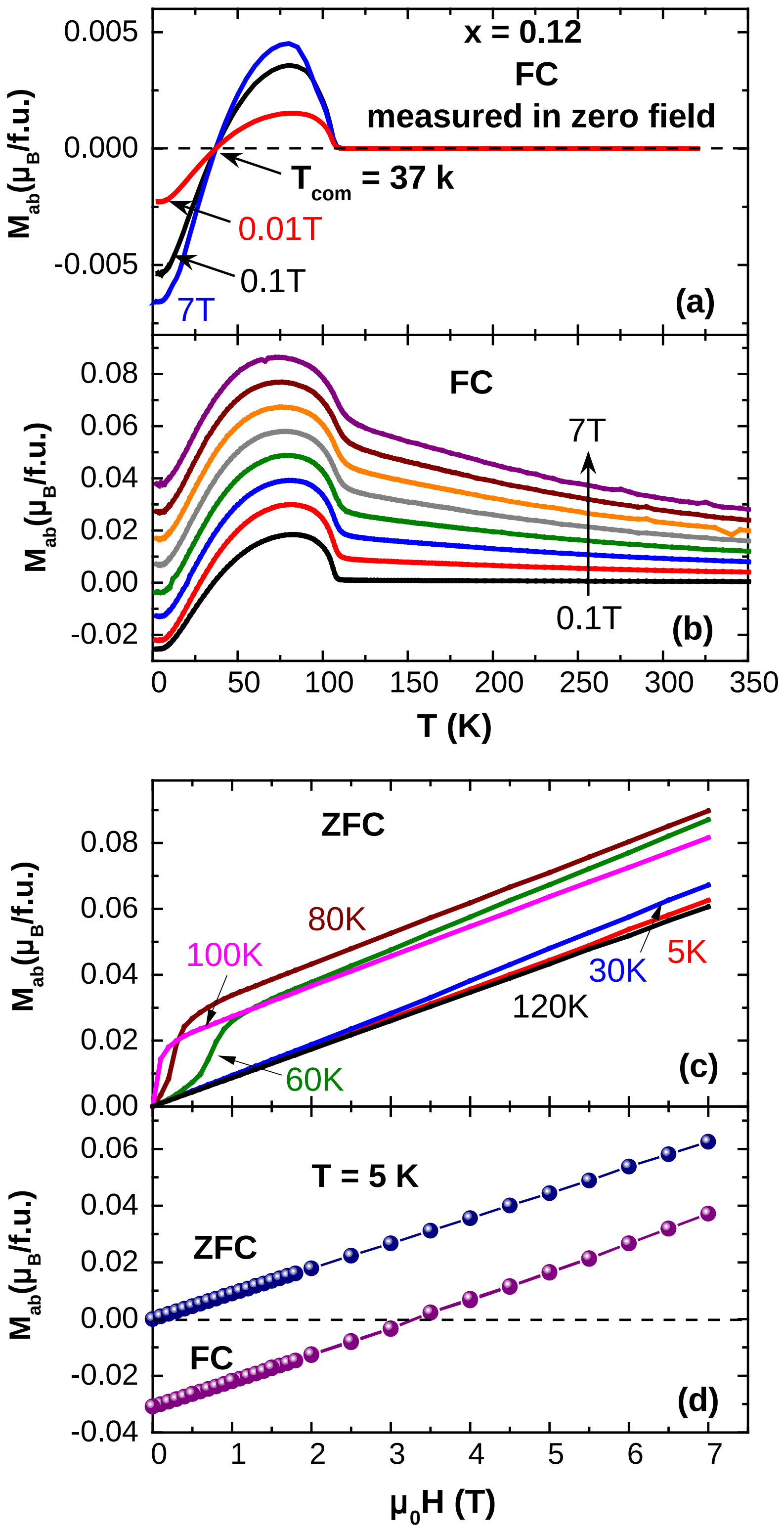

Fig. 2

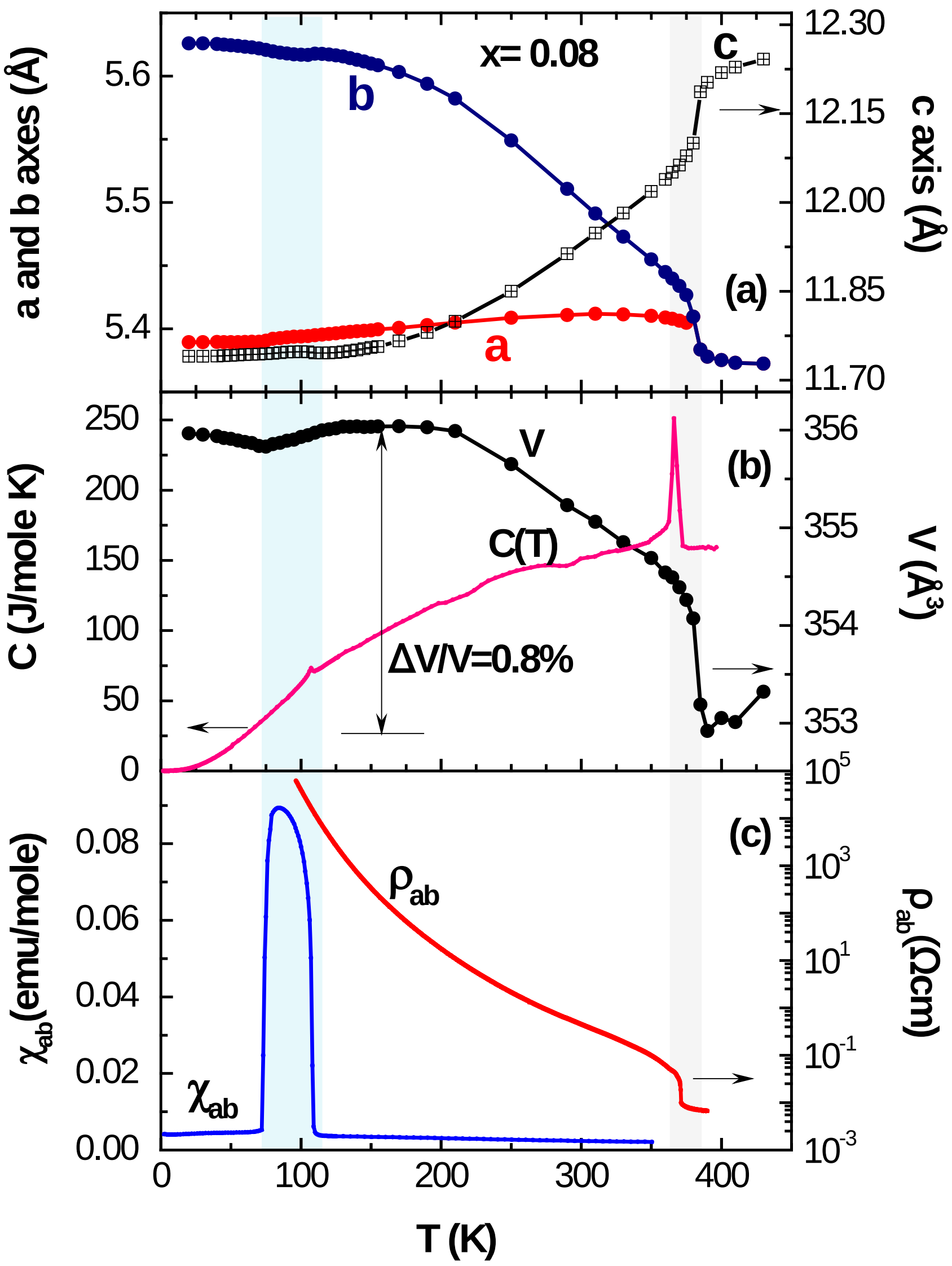

Fig. 3

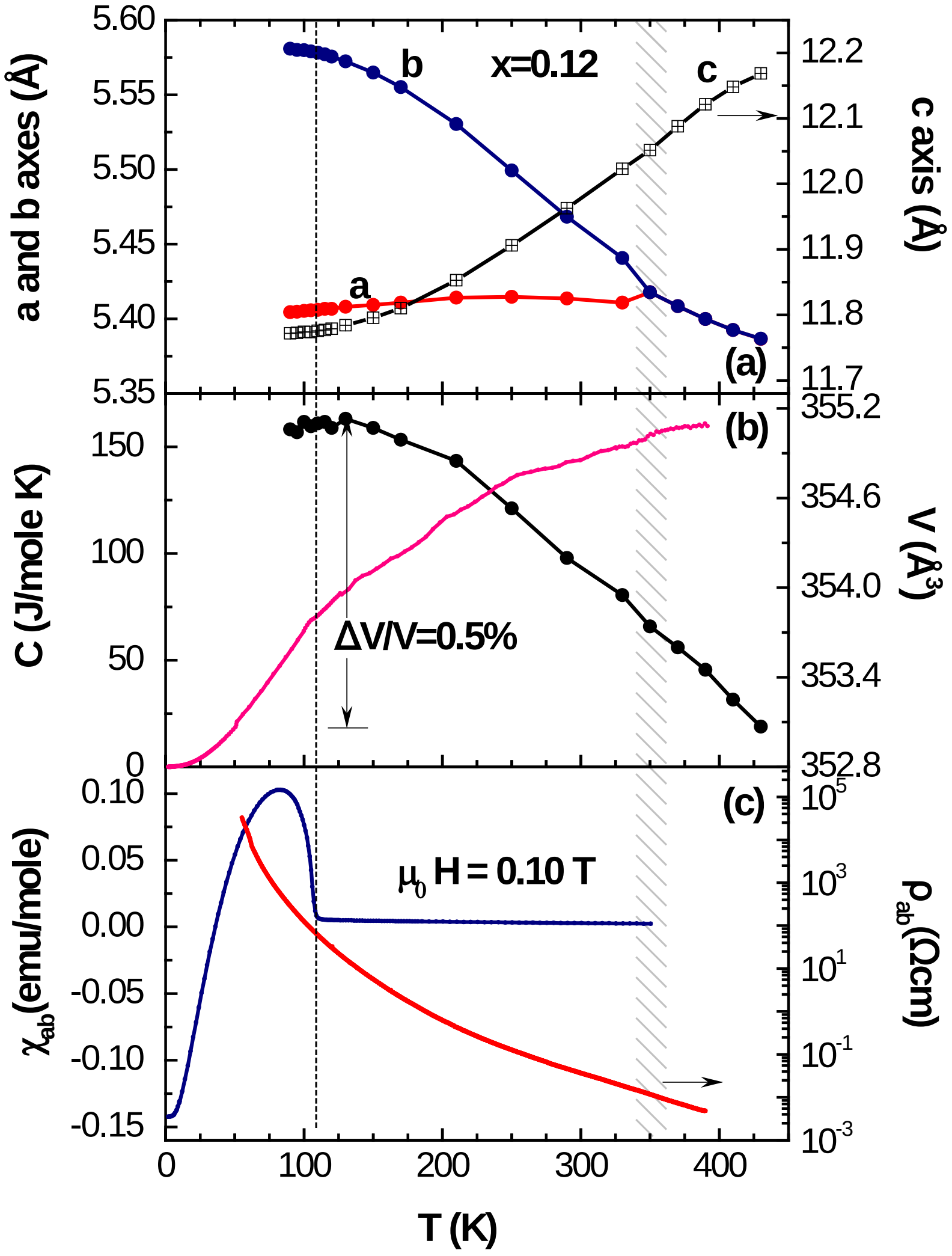

Fig. 4

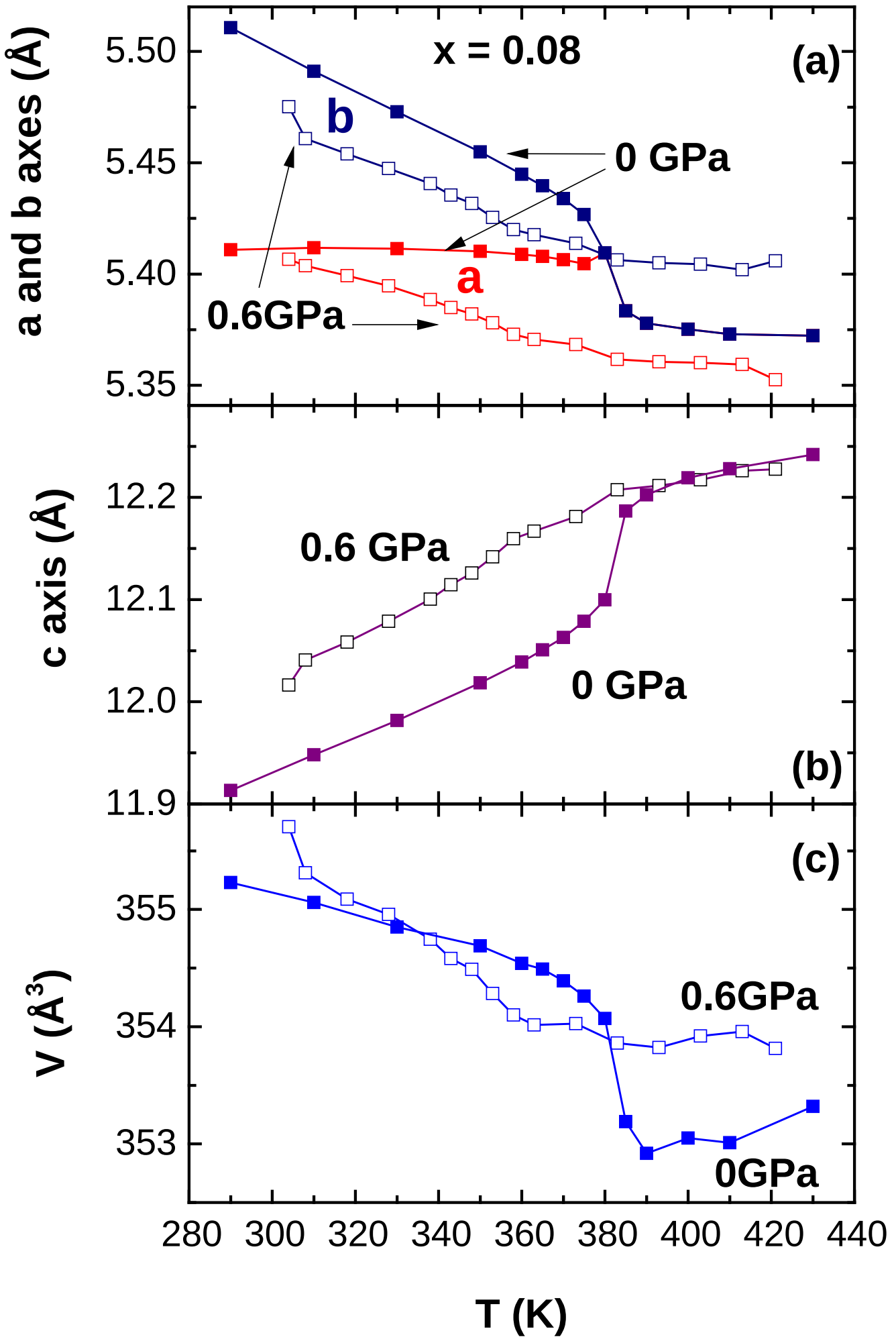

a and b axes (Å)
5.50
5.45
5.40
5.35
x = 0.08
(a)
b
0 GPa
a
0.6GPa
c axis (Å)
12.2
12.1
12.0
11.9
0.6 GPa
0 GPa
(b)
V (Å$^3$)
355
354
353
(c)
0.6GPa
0GPa
280 300 320 340 360 380 400 420 440
T (K)

Fig. 5

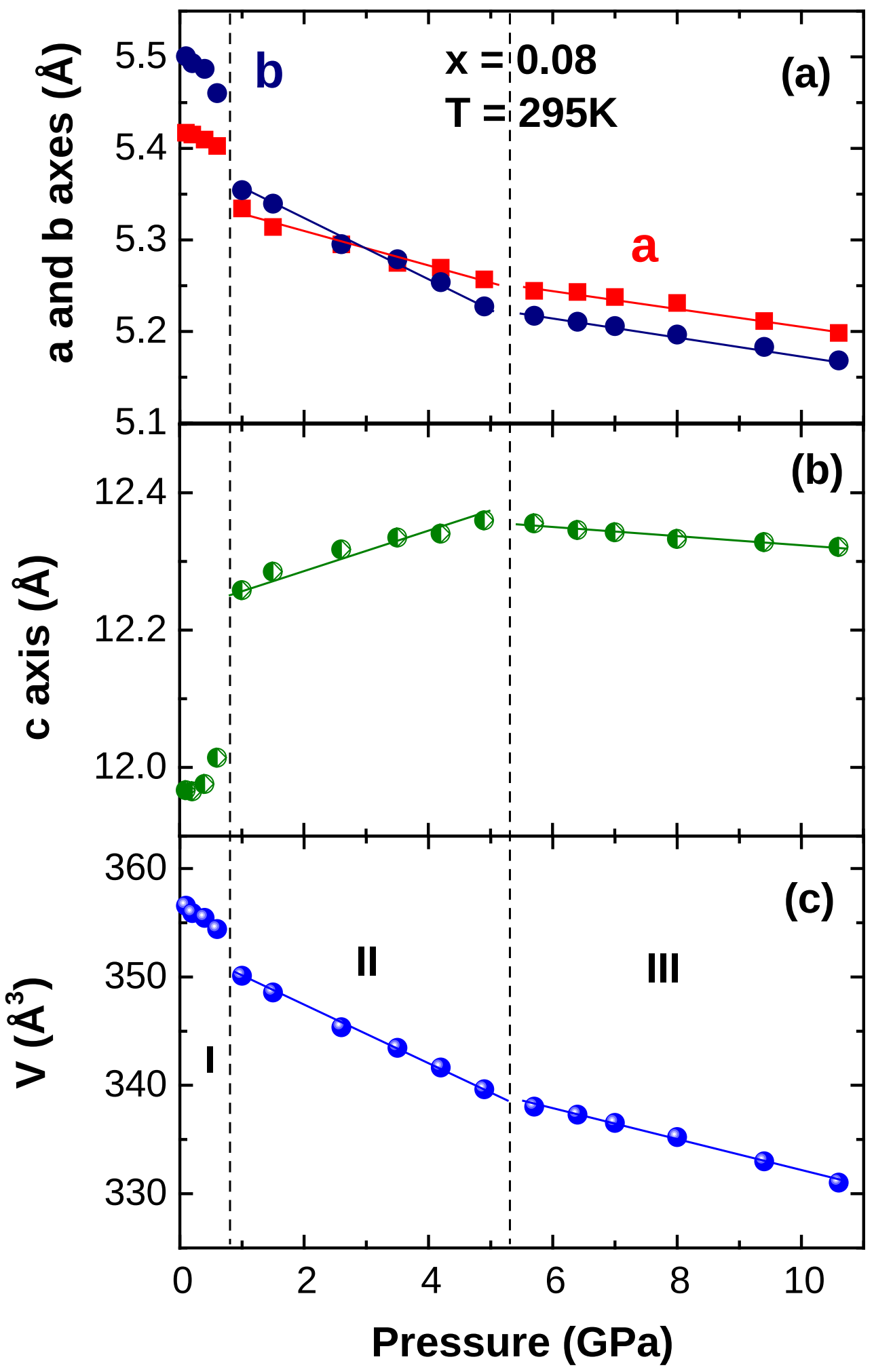

Fig. 6